\documentclass[traditabstract]{aa}

\usepackage{graphicx}
\usepackage{natbib}
\usepackage{lscape}
\usepackage{supertabular}
\usepackage{xcolor}

\newcommand{\vsini}{$v \sin{i}$}

\newcommand{\rp}[0]{R$^{\prime}_{HK}$\,}
\newcommand{\rplus}[0]{R$^{+}_{HK}$\,}
\newcommand{\cahk}[0]{\ion{Ca}{ii}~H\&K}
\newcommand{\mws}[0]{\mbox{S$_{\rm MWO}$}}

\newcommand{\cairt}[0]{\ion{Ca}{ii}~IRT\,}
\newcommand{\lxlbol}[0]{log(L$_{\rm X}$/L$_{\rm bol}$)\,}

\usepackage{txfonts}

\begin{document}

\title{The corona -- chromosphere connection studied with simultaneous eROSITA and TIGRE observations
\thanks{Full Table 3 is only available in electronic form at the CDS via anonymous ftp to cdsarc.u-strasbg.fr}
}

\author{B. Fuhrmeister\inst{\ref{inst1}} \and S. Czesla\inst{\ref{inst1}}   
  \and J. Robrade \inst{\ref{inst1}}
  \and J.~N. Gonz\'alez-P\'erez\inst{\ref{inst1}}
  \and C. Schneider\inst{\ref{inst1}} 
  \and M. Mittag \inst{\ref{inst1}} 
  \and J. H. M. M. Schmitt\inst{\ref{inst1}}
}      

\institute{Hamburger Sternwarte, Universit\"at Hamburg, Gojenbergsweg 112, D-21029 Hamburg, Germany\\
  \email{bfuhrmeister@hs.uni-hamburg.de}\label{inst1}
       }

\date{Received dd/mm/2020; accepted dd/mm/2020}


\abstract{Stellar activity manifests itself in a variety of different phenomena, some of which we
can  measure as activity tracers from different atmospheric layers of the star, typically
at different wavelengths.   Stellar activity is furthermore inherently time variable,
therefore simultaneous measurements are necessary to study the correlation between
different activity indicators.   In this study we compare X-ray fluxes measured within 
the first all-sky survey conducted by the extended ROentgen Survey with an Imaging
Telescope Array (eROSITA) instrument on board the Spectrum-Roentgen-Gamma (SRG) observatory 
to \cahk\, excess flux measurements \rplus, using 
observations made with the robotic TIGRE telescope.  We created the largest sample of simultaneous X-ray and spectroscopic \cahk\, observations of late-type stars obtained so far, and in addition,
previous measurements of \cahk\, for all sample stars were obtained.
We find the expected correlation between our \lxlbol to log(\rplus)  measurements,
but when the whole stellar ensemble is considered, the correlation between
coronal and chromospheric activity indicators does not improve when the
simultaneously measured data are used. A more detailed analysis shows that the
correlation of \lxlbol to log(\rplus)  measurements of the
pseudo-simultaneous data still has a high probability of being better than that of a random set of
non-simultaneous
measurements with a long time baseline between the observations.
Cyclic variations on longer timescales
are therefore far more important for the activity flux-flux relations than short-term variations in the
form of rotational modulation or flares, regarding the addition of ``noise'' to the activity
flux-flux correlations. Finally, regarding the question of predictability of necessarily
space-based
\lxlbol measurements by using ground-based chromospheric indices,
we present a relation for estimating \lxlbol from \rplus\, values
and show that the expected error in the calculated minus observed (C-O) \lxlbol values is 0.35 dex. }

\titlerunning{Chromospheric--coronal connection by eROSITA and TIGRE}
\authorrunning{B. Fuhrmeister et~al.}
\maketitle


\section{Introduction}

Magnetic stellar activity manifests itself in a zoo of phenomena, originating
in various layers of the stellar atmosphere. This is observable throughout the electromagnetic
spectrum.  In the optical, photometric variability \citep[e.g.][]{SM16} is observed that
is attributed to star spots, which are located in the stellar photospheres and  
rotate in and out of the field of view.
Activity phenomena of the chromosphere and transition region are frequently studied
using several optical and ultraviolet spectral lines \citep{Baliunas1996, Wright2004}.
Finally, the tenuous and hot coronal gas can be observed at X-ray wavelengths \citep{Guedel2004,Robrade2005}.

All these different tracers of magnetic activity are ultimately 
thought to be powered by a stellar dynamo, operating in the interiors of stars with
outer convection zones. 
Chromospheric emission lines additionally show
a so-called basal level that may be caused
by energy dissipation of acoustic waves in the chromosphere and perhaps the transition region 
\citep{Rutten1991, Wedemeyer2004}. A comparable basal level has not been observed in
X-rays \citep{Stepien1989,Schrijver1992}. Although a minimum flux 
observed by \citet{Schmitt1997} also exists, it is more than two magnitudes higher than
the expected value from acoustic heating \citep{Stepien1989}. Instead, it
coincides with the X-ray flux level from solar coronal holes, which represents
the minimum observed flux from the solar corona.

Optical spectral and photometric activity tracers have been studied in
time-series observations of individual stars, which revealed activity cycles
\citep{Hempelmann2006}, stellar rotation periods \citep{SM16, Patrick, rotation}, and  
flares \citep{CNLeoflare}, for instance. These studies have allowed analysing the
relation between different activity tracers, and they showed that the tracers
tend to be well correlated.
However, many authors also find that this is not necessarily true for all activity
tracers all the time. For example,
\citet{GomesdaSilva} found in a study of M~dwarfs
that the emission in the H$\alpha$ and the \cahk\ lines
need not correlate in lower-activity stars, while this is the case for more active
stars. They therefore recommended using either the \cahk\ lines or the 
\ion{Na}{i} D lines as a proxy for the activity level in M~dwarfs.
Nonetheless,
there may be good correlations between the H$\alpha$ and
the \cahk\ lines for individual stars \citep[e.g.][]{Robertson}.
Considering earlier type stars as well, 
\citet{Diaz2007} concluded from a study of F6 to M5.5 stars that \ion{Na}{i} D can only be used
as activity indicators for the most active stars.

Using an ensemble of diagnostics, activity throughout the stellar atmosphere can be
studied. For example, \citet{Walkowicz2009} analysed the correlation between the emission in
the \cahk, H$\alpha$, and \ion{Mg}{ii} ultraviolet lines and the X-ray flux in a sample of M3 stars.
These indicators trace layers of the stellar atmosphere with increasing height. \citet{Walkowicz2009}
found a good  correlation between fluxes originating in the lower and upper chromosphere
as traced by the \cahk\ and H$\alpha$ lines, again
only for active stars. In stars with a high or intermediate activity level, power-law 
relations were found between
the X-ray coronal and optical chromospheric line tracers. 

Comparisons of the X-ray flux and the \cahk\ excess emission were also performed for more solar-type stars \citep{Schrijver1983, Mittag2018}.
The power-law relation between the chromospheric \cahk\ emission and the coronal X-ray flux
holds over four orders of magnitude in X-ray flux; the Sun also follows this relation during
its activity cycle.
In a similar study,
\citet{MartinezArnaiz2011} deduced power-law dependences between several
chromospheric and X-ray indicators using a spectral subtraction technique that eliminates photospheric and basal contributions in their simultaneous
chromospheric data. They derived flux--flux relations for the H$\alpha$,
\cahk, and \cairt\ lines and the X-ray flux using 298 F- to M-type stars, including some younger
stars. \citet{He2019} found a similar correlation between \lxlbol\
and log(L$_{\mathrm{H\alpha}}$/L$_{\mathrm{bol}}$) for 484 F- to M-type stars.  

The correlation between chromospheric and coronal fluxes extends over many orders of magnitude and 
hence is quite striking, but it shows considerable scatter, the reasons for which are unclear.
A possible explanation clearly is that the vast majority of data sets does not provide simultaneous 
chromospheric and coronal measurements, and the chromosphere and corona of the same
star are compared, but at very different times and therefore at possibly 
different activity levels.  In an attempt to address this issue,
\citet{Schrijver1992} studied a small sample of stars for which nearly simultaneous measurements 
 of the fluxes in the \ion{Mg}{ii} h\&k lines (with the IUE satellite), of the broad-band soft X-ray flux
(with the EXOSAT satellite), and the Ca~II H\&K core emission as determined from Mount Wilson S-indices were available.  
\citet{Schrijver1992} reported that the scatter is reduced when near-simultaneous data are used and that
the remaining scatter can be entirely  attributed to
uncertainties in the instrumental calibration and flux conversion factors.

To increase the sample size and improve the temporal alignment,
we used our robotic TIGRE telescope to obtain the largest sample 
of (pseudo-)simultaneous, high-resolution
optical spectra of stars observed at X-ray wavelengths so far in the framework of the
first all-sky survey by the extended ROentgen Survey with an Imaging
Telescope Array (eROSITA). Because we have non-simultaneous data as well, we can
analyse the effect of the simultaneity on the power-law distribution connecting the X-ray flux
and chromospheric tracers. We wish to study in particular whether  
part of the scatter in this relation is due to non-simultaneous data and explore
in this context how well \lxlbol\, can be predicted by
chromospheric indicators measured from the ground.

\section{Stellar sample and observations}
\label{sec:obs}
In this section we first describe our stellar sample, followed by an overview of our
X-ray eROSITA and optical TIGRE observations. We also provide a description of
the applied scheduling procedures to obtain a pseudo-simultaneous data set 
and characterise these data. 
Finally, we introduce the chromospheric activity indicator \rplus.

\subsection{Stellar sample}

To accomplish our goals, we need a sample of stars that is as equally distributed in spectral type
and activity level as possible, with a
preferably small fraction of low-activity stars without  X-ray detection by eROSITA
and bright enough to be observed by the TIGRE telescope;
a representative volume-limited sample is therefore not the best choice.
We therefore constructed a sample of 404 stars with spectral types between M and F
indicated by a colour constraint\footnote{We adopted B-V values from SIMBAD \citep{simbad} for all stars but HD~97503,
where the Hipparcos main catalogue was used \citep{Hipparcos}.}
of 0.24 $\leq$ B - V $\leq$ 1.5 mag.
(see Table~\ref{samplestars}). From these, we selected all optically bright (V $\leq 9.5$~mag) stars,
which must additionally be located
in the western galactic hemisphere at galactic longitudes between 180$^{\circ}$ and
360$^{\circ}$, where eROSITA data is available to us. 
 Moreover, we applied an individual distance 
cut for  spectral type groups  provided in Tab.~\ref{samplestars}.
The distance cut was larger for stars with a  detection by the ROSAT satellite 
(count rate in excess of
0.01~cts/s) compared to stars without. Resolvable binaries were excluded from our sample.
However, we 
included 48 binary candidates covering spectroscopic binaries or RS CVn systems,
which are unresolved by eROSITA and TIGRE.
We discuss these stars further in Sect. \ref{Sec:corr} and
Sect. \ref{secspectype}.  

Thus 
our sample is a compromise between a  volume-limited sample and a
more active sample in which we expect more eROSITA detections. The sample has a slight
bias on solar-type and earlier stars because the fainter later stars require far longer
observation times in the optical. 

\begin{table}
\caption{\label{samplestars} Basic properties of the sample stars. }
\footnotesize
\begin{tabular}[h!]{lccc}
\hline
\hline
\noalign{\smallskip}

B-V & spectral types & d$_{\mathrm{cut}}$ w/wo 2RXS   & number of  \\
$[$mag$]$ &              &  detection [pc] & stars\\
\noalign{\smallskip}
\hline
\noalign{\smallskip}
$<$ 0.44 & F0 -- F5 & 50/20 & 65 \\
0.44 -- 0.58 & F5 -- F9 & 50/20 & 85 \\
0.58 -- 0.77 & F9 -- G9 & 50/30 & 140 \\
0.77 -- 1.25 & G9 -- K6 & 30/20 & 89 \\
$>$ 1.25     & K6 -- M  & 30/20 & 25 \\
\noalign{\smallskip}
\hline

\end{tabular}
\normalsize
\end{table}

\subsection{eROSITA data}\label{erosita}

The eROSITA X-ray telescope is the main instrument on board the Spectrum-Roentgen-Gamma satellite
\citep[SRG,][]{Predehl2020}. The SRG is carrying out a total
of eight all-sky surveys in the $0.2-8$~keV range and performs the first ever systematic 
all-sky survey in the $2.4-8$~keV X-ray band. Thus, the eROSITA survey complements the 
ROSAT all-sky survey (RASS) carried out in 1990 in the $0.1-2.4$~keV energy range, and significantly
surpasses it in terms of sensitivity, spatial, and spectral resolution.

For our study we only used results of the first of the eight eROSITA all-sky surveys, using the processing
results of the eSASS pipeline (currently version 946); a detailed description
of this software and the algorithms we used is given by \citet{Brunner2021} (in preparation).
Briefly,  the eSASS system performs source detection and creates X-ray catalogues for
each sky tile with a size of 3.6$^{\circ}$ x 3.6$^{\circ}$ in which neighbouring sky tiles overlap.
The X-ray sources in the individual tiles are then merged into one large catalogue.\ Duplications
in the overlap regions were removed and astrometric corrections were applied. These catalogue count rates are
fiducial count rates that would be obtained if the source in question were observed by all seven
eROSITA telescopes on-axis; we note that during the survey, the off-axis angle changes all the time,
and furthermore, not all seven telescopes may deliver useful data at all times due to calibration
activities or malfunction.   The software performs source detection in the energy bands 0.2~keV~-~0.6~keV,
0.6~keV~-~2.3~keV, and 2.3~keV~-~5.0~keV and also considers a total band 0.2~keV~-~5.0~keV.  Because
stellar coronae tend to be rather soft X-ray emitters, 
most of our sample stars have no counts, or counts that agree with zero counts in the highest-energy band. We therefore only used the energy bands 0.2~keV~-~0.6~keV and
0.6~keV~-~2.3~keV for our analysis.  

Furthermore, the measured count rates in the two lower-energy bands
show a high degree of correlation.  We therefore used the summed count rates in the 0.2 - 2.3~keV band and applied an energy conversion factor (ECF) of
0.9 $\times$ 10$^{-12}$ erg cm$^{-2}$ cnt$^{-1}$  to convert measured count rates into inferred energy fluxes. We note that
this ECF is appropriate for a thermal plasma emission with a temperature of 1~keV and an absorption column
of $3\cdot 10^{19}$ cm$^{-2}$. No attempt was made to apply individual conversion factors to individual sources
based on their X-ray spectra because many of our sources are too weak to meaningfully fit a coronal spectral model. We estimate that the errors of our preliminary calculations lead to flux
errors of about 10 to 20\%. This is supported by simulations with thermal models 
of a single-temperature coronal plasma and the eROSITA response function, which show that the ECF varies by
about 10 \% for coronal temperature changes in the range between 0.2~keV to 1.0~keV.

To further bolster our error estimates, 
we compared the tabulated fluxes to ROSAT
fluxes obtained with the hardness-ratio-dependent conversion factor
given by \citet{Schmitt1995}, which is based on
coronal spectral models. The ROSAT fluxes calculated in this way are
about 25 \% higher than their eROSITA counterparts. This can mostly be attributed to
the softer ROSAT energy band, which leads to systematically higher fluxes by 15 \% using standard
coronal models. 

Proceeding in this fashion,
we identified X-ray counterparts for 343 of our sample stars in eRASS1 and
320 in the RASS catalogue (2RXS) by \citet{Boller2016}, see Table~\ref{xraysources}.
We used a search radius of 18 arcsec for eRASS1 and a search radius of 40 arcsec for
the ROSAT catalogue. We estimate our completeness to be more than 95 \%, with a marginal
number of by-chance-detections.

Thirty-eight stars have  X-ray detections in eRASS1 but not in 2RXS. About
half of them have low count rates at the detection threshold of eRASS1, demonstrating
the higher sensitivity of eROSITA compared with ROSAT. \citet{Boller2016} stated a detection
limit in 2RXS of 10$^{-13}$~erg\,cm$^{-2}$\,s$^{-1}$. The X-ray emission of 18 of these 38 stars without a ROSAT
detection is below that value. 
Another 7 stars without a ROSAT detection
have flux values below 2$\cdot$10$^{-13}$ erg\,cm$^{-2}$\,s$^{-1}$. Of the remaining stars, 3
have counts in the hardest band and may therefore have undergone flares during the eROSITA observation.
The rest may be explained either by long-term activity variation (cycles) or spurious detections.



\begin{figure}
\begin{center}
\includegraphics[width=0.5\textwidth, clip]{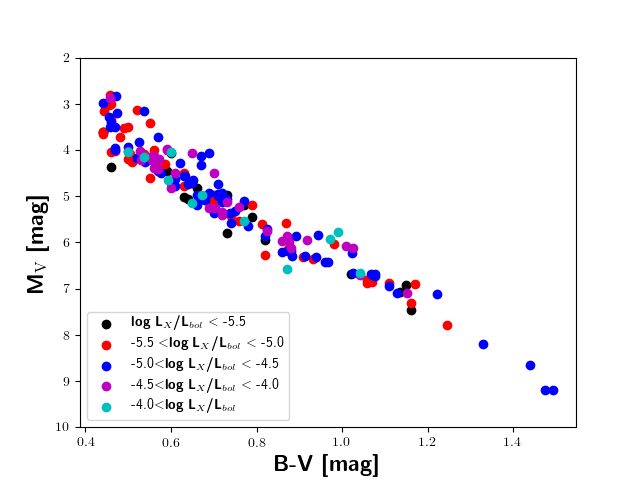}\\
\caption{\label{hrdiagram} HR diagram of a sub-sample of our stars defined
as a valid sample in Sect. \ref{characterization} that also shows the X-ray
properties of these stars. 
}
\end{center}
\end{figure}

Finally, we converted the measured X-ray fluxes into X-ray luminosities using parallaxes from {\em Gaia} DR2 \citep{Gaia} and
Hipparcos \citep{Hipparcos} and calculated the bolometric luminosities of our stars
by interpolating the values given by \citet{Mamajek}\footnote{\tt http://www.pas.rochester.edu/\~emamajek/EEM\_dwarf\\ \_UBVIJHK\_colors\_Teff.txt}.
To provide an overview of the properties of our stellar sample, we show
in Fig. \ref{hrdiagram} a Hertzsprung-Russell (HR) diagram of a
sub-sample of our program stars, defined in Sect.~\ref{characterization} as a valid
sample (all stars with an X-ray counterpart, but excluding earlier-type stars, sub-giants, and noisy
data).
Figure~\ref{hrdiagram} shows the smooth distribution of stars in B-V as well
as their X-ray properties characterised by 
their eROSITA-measured \lxlbol \ values.

\subsection{TIGRE data}

We used our robotic TIGRE telescope to obtain high-resolution ($R\approx 20\,000$) optical spectra
of our sample stars. TIGRE is a 1.2~m telescope located in central Mexico at the La Luz Observatory 
of the University of Guanajuato. The telescope is fibre-coupled to an \'echelle spectrograph with a 
wavelength coverage from 3800~\AA \,to 8800~\AA.  From the 2D \'echelle images, 1D science 
spectra are obtained via a reduction pipeline; a detailed description of the TIGRE facility is given by
\citet{Schmitt2014}.

The spectral range of TIGRE is divided into two arms. The blue arm
covers the spectral range shortwards of $5800$~\AA,\ and the red arm covers the range longwards of $5900$~\AA,\ 
extending up to about $8800$~\AA . 
Thus, the blue arm covers the \cahk\ lines, which we focus upon in this paper, while the red arm
contains the H$\alpha$ and the \cairt\ lines, which also provide chromospheric diagnostics.
Along with the 1D science spectra, the reduction pipeline also determines activity indices, and
the instrumental \cahk\ index can be converted into the Mount Wilson Observatory S-index (\mws) using
the relation by \citet{Mittag2016} to ensure comparability with other measurements.

\subsubsection{Characterisation of the pseudo-simultaneous data}\label{characterization}

The X-ray measurements span some time because eROSITA scans the sky  along great circles with a 
scan period of 4~hours; thus any source stays within the eROSITA field of view for at most 40 seconds per scan.
The scanned great circles move along the plane of the ecliptic with a (somewhat variable) rate of about
1$^{\circ}$ per day, so that a source near the plane of the ecliptic is visited six times within 24 hours,
while sources at higher ecliptic latitude are observed longer. 
We define the middle of the eROSITA observations as the average
of time between the first scan and last scan (even if at that time the star was actually not being scanned). To provide
a few numbers, the longest scan period of eROSITA among our stars is 6.5 days, and about 20 stars
have scan elapsed times longer than two days. For the purposes of this paper, it is essential to
obtain spectral observations pseudo-simultaneously with
the eROSITA X-ray measurements. In the context of this study, we consider all optical spectra
taken within three~days before or after the start or stop of the X-ray observation as pseudo-simultaneous.

Next, the great circles scanned by eROSITA are more or less
90$^{\circ}$ away from the Sun, and therefore eROSITA can never observe in the anti-Sun direction that is normally favoured
by optical observers.  As a consequence, 
TIGRE observations have to take place either shortly after sunset or just prior to sunrise, and the
allowed observing window can be as short as 24 hours.   Scheduling these (pseudo-)simultaneous 
observations therefore is a challenge and certainly a task ideal for robotic telescopes.  For reference,
we also obtained non-simultaneous spectra of our program stars.

The TIGRE exposure time is set to obtain a signal-to-noise ratio (S/N) of 100 at 6000~\AA, but
for spectra to be usable for Ca~II H\&K studies, we demand a posteriori a mean S/N greater than 20 in the blue arm, 
which we found necessary to compute a meaningful \cahk\ index. 
During the eRASS1 data taking,  our pseudo-simultaneous TIGRE observation program therefore led to
550 usable spectra for a total of 312 stars, see Table \ref{xraysources}.

Moreover, when we also consider the non-simultaneous observations of the sample stars,
 we have 9 or more spectra for the majority of stars, and for
a sub-sample of 15 stars, we even have 100 or more spectra. Unfortunately,
 for 10 faint and very late-type stars, no spectra were obtained with usable \cahk\ data. For another 15 of the faintest stars, we have only one or two usable spectra.

\begin{figure}
\begin{center}
\includegraphics[width=0.5\textwidth, clip]{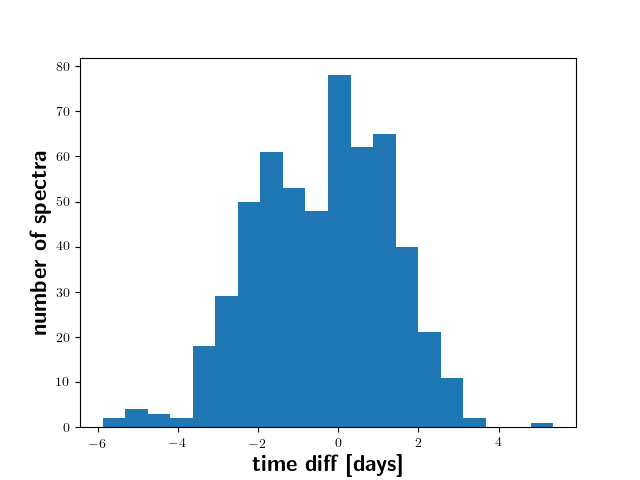}\\
\caption{\label{timediff} Number of spectra taken by TIGRE for a certain offset between
  the middle of the eROSITA observation and the time of the TIGRE observations for pseudo-simultaneous data.
}
\end{center}
\end{figure}

In Fig.~\ref{timediff} we show the distribution of the temporal offset between the TIGRE observations and the time centre of the eROSITA
observation for the usable pseudo-simultaneous
TIGRE observations. The time centre of the eROSITA observations is defined as the mid-time
between the first and last eROSITA scan as explained above.
Two peaks can clearly be distinguished in the distribution, separated by about two days,
but Fig.~\ref{timediff} shows that our goal of reaching pseudo-simultaneous observations
has been accomplished quite well. 

The number of stars with eROSITA X-ray measurements and pseudo-simultaneous TIGRE data that lie in
the colour range B-V $>$ 0.44 mag (to ensure that the conversion relation to \rp and \rplus 
values is valid; cf. Sec.~\ref{sec_conv}) is 203. From this sample we excluded 
 20 stars that
are outliers in our HR diagram (most of them are probably sub-giants) or are outliers in \rplus, which normally is found for stars
with very low surface gravities  or peculiar metallicity values.
This selection leaves us with 183 stars, which we call the valid sample in the following.

\begin{table}
\caption{\label{xraysources} Observational properties. }
\footnotesize
\begin{tabular}[h!]{lcccc}
\hline
\hline
\noalign{\smallskip}

               & number of stars \\

\noalign{\smallskip}
\hline
\noalign{\smallskip}
2RXS detected& 320  \\
eRASS1 detected & 343\\ 
eRASS1 or 2RXS detected & 358 \\
eRASS1 and 2RXS detected & 305 \\
eRASS1 but not 2RXS detected & 38\\
2RXS but not eRASS1 detected & 15\\
pseudo-simultan. \cahk & 312 \\
pseudo-simultan. \cahk + eRASS1 detected & 258 \\
average \cahk & 394 \\
\noalign{\smallskip}
\hline

\end{tabular}
\normalsize
\end{table} 

\subsubsection{Conversion into \rp and \rplus}
\label{sec_conv}

The so-called Mount Wilson S-index (\mws) as defined by \citet{Wilson1978} is a purely instrumental index that must be
calibrated to obtain physical quantities and allow comparisons between
stars of different spectral type.  Furthermore, it
contains both chromospheric and photospheric non-activity-related contributions,
so that we have to remove the latter,
which depend on stellar effective temperature or colour, respectively.

Comparisons between stars with different spectral types  become possible by the introduction
of the \rp index (empirically defined by \citet{Hartmann1984, Noyes1984}). This is
the ratio of the chromospheric \cahk\
line fluxes and the bolometric flux.  \citet{Mittag2013} revisited the S-index calibration, taking
advantage of synthetic PHOENIX models  \citep{Hauschildt1999}, and introduced an additional index,
\rplus, that differs from \rp by the subtraction of a basal activity level
from the \cahk\ line flux.
This basal activity level is thought to be caused by heating by
acoustic waves that are produced by turbulent convective motions and  deposit their mechanical
energy in the chromosphere \citep{Schrijver1995, Wedemeyer2004}. Although previous
studies did not introduce the term \rplus\ , the technique of identifying and eventually subtracting this
non-magnetic heating from the line flux has had a long tradition (see e.g. \citet{Schrijver1987, Mathioudakis1992}).

We first converted S$_{\rm TIGRE}$ into the \mws\ index using the relation by
\citet{Mittag2016} and then into \rp using two relations, first, the ``classical'' relation given
by \citet{Noyes1984}, and, second, the more recent relation published by \citet{Mittag2013}. The latter also allows conversion
into the \rplus index.
The relations by \citet{Noyes1984} and \citet{Mittag2013}
are only defined for stars within the colour ranges $0.44 < {\rm B}-{\rm V} < 0.9$ and
$0.44 < {\rm B}-{\rm V} < 1.6$, respectively.
Unfortunately, 65 earlier-type program stars (see Table~\ref{samplestars})
are outside either range. We excluded them from the conversion for the time being.
Late-type stars with B$-$V colour larger than $0.9$, that is, beyond the range of validity
defined by \citet{Noyes1984}, were not discarded for
comparison purposes, with the relation extrapolated to these B$-$V values.

\begin{figure}
\begin{center}
\includegraphics[width=0.5\textwidth, clip]{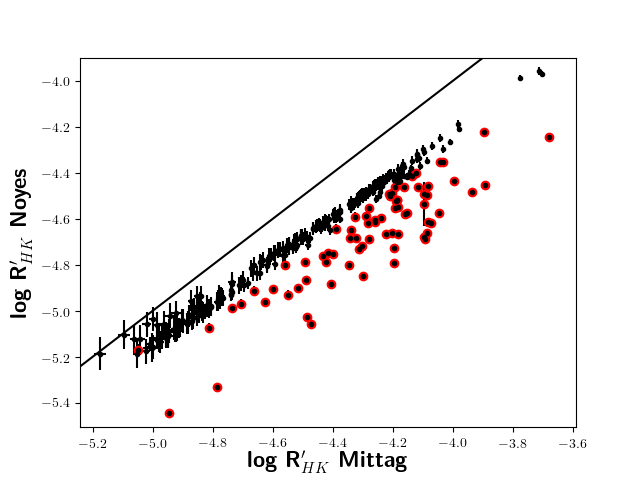}\\
\caption{\label{NoyesMarco} \rp values computed using the relation given by \citet{Noyes1984}
  compared to the relation given by \citet{Mittag2013}. The red dots mark the stars with B$-$V > 0.9. This  extends the relation given by \citet{Noyes1984} beyond its definition region.
}
\end{center}
\end{figure}

We compare the \citet{Noyes1984} and \citet{Mittag2013} conversions into \rp
in Fig.~\ref{NoyesMarco}. The relation proposed by \citet{Mittag2013}
leads to systematically higher \rp values, and the difference slightly grows toward
the most active stars.
Because the two relations are derived from a different ansatz, some discrepancy may be expected.
With the available data, it cannot be decided which relation is better. We nevertheless 
used the relation of \citet{Mittag2013} because we intend to use \rplus.
Fig.~\ref{NoyesMarco} also demonstrates that an extrapolation  of the relation
given by \citet{Noyes1984} toward larger B$-$V colour seems inappropriate.

\section{Sample properties}

In Table~\ref{measurements} we summarise the key properties of the sample data.
Specifically, we list \lxlbol\, values computed from the eRASS1 X-ray flux,
the median of the \rplus  values of all spectra, and those of the pseudo-simultaneous spectra. Moreover, we
list the original measurements of the median of S$_{Tigre}$ of all spectra and of the pseudo-simultaneous 
spectra. S$_{Tigre}$ can be linearly scaled to S$_{MWO}$  using the conversion given
in \citet{Mittag2016}. 
Finally, we state the time difference between the middle of the eROSITA observation window and
the nearest TIGRE pseudo-simultaneous
observation.

We further list an observed and a computed rotation period, the derivation of which is described
in Sect.~\ref{rotation}.
The observed
period is the median value of all observational results given in the catalogues listed as
footnote to Table~\ref{measurements}. Moreover, we specify whether the star is in the valid
sub-sample as defined in Sect. \ref{characterization}.
We show the first ten rows in Table~\ref{measurements} for orientation,
and the whole table is available electronically at CDS.

\begin{sidewaystable}
\caption{\label{measurements} Measured properties of the sample stars$^{a}$. }
\footnotesize
\begin{tabular}[h!]{lccccccccccccccc}
\hline
\hline
\noalign{\smallskip}

Name       & B-V & median   & err     & median      & err    & median & err & median & err &median & err &        T$_{diff}$ & P$_{rot}$ & P$_{rot}$ &valid \\
           &     & \lxlbol &\lxlbol &S$_{Tigre}$ & S$_{Tigre}$ & S$_{Tigre;sim}$ & S$_{Tigre;sim}$ & log(\rplus) & log(\rplus)& log(\rplus$_{sim}$) & log(\rplus$_{sim}$) &                 &  lit & comp.  &   \\
           & [mag]&    &   & [10$^{-3}$\AA] & [10$^{-3}$\AA] & [10$^{-3}$\AA] &[10$^{-3}$\AA]   & & &  &  & [days] & [days] & [days] & \\
\noalign{\smallskip}
\hline
\noalign{\smallskip}
\footnotesize
        HD61421  &    0.42   &    -5.827  &     0.012    &   6.339  &     0.038    &   6.306  &     0.010   &     0.000  &      0.000      &  0.000   &     0.000   &     0.000  &       3.000  &         inf   &    0\\
        HD30652  &    0.44   &    -5.192   &    0.020  &     7.927  &     0.034 &      7.169    &   0.078     &  -4.778   &     0.006    &   -4.926   &     0.018   &    -0.574      &   0.000     &    6.456   &    1\\
        HD48737  &    0.43     &  -4.988    &   0.025    &   8.823     &  0.043    &   8.788 &      0.045   &     0.000     &   0.000     &   0.000    &    0.000 &      -0.274  &       0.000       &    inf    &   0\\
        HD23249   &   0.92    &   -5.235   &    0.041  &     5.191   &    0.055  &     5.115   &    0.046   &    -5.751   &     0.037   &    -5.806    &    0.036 &      -5.888     &   71.000     &   64.310  &     0\\
        HD38393  &    0.47    &   -6.464  &     0.056   &    5.444   &    0.044    &   0.000    &   0.000   &    -5.842   &     0.082   &     0.000  &      0.000 &       0.000 &       21.400   &     12.832  &     0\\
        HD102870  &   0.55  &     -5.447  &     0.027 &      5.988   &    0.040   &    6.039   &    0.035  &     -5.254    &    0.017   &    -5.233  &      0.014   &    -1.367     &    0.000    &    17.091   &    1\\
        HD40136   &   0.33  &     -5.787  &     0.030    &   9.483    &   0.072   &    0.000 &      0.000   &     0.000   &     0.000     &   0.000   &     0.000  &      0.000    &     0.000    &       inf    &   0\\
        HD22049  &    0.88   &    -4.963  &     0.012   &    20.790  &     0.116  &     20.791  &     0.082  &     -4.306 &       0.003   &    -4.306   &     0.002   &    -3.430      &  11.840   &     11.629   &    1\\
        HD20010   &   0.51    &   -4.374    &   0.007    &   4.832     &  0.059   &    4.814   &    0.058   &     0.000   &     0.000  &      0.000    &    0.000   &    -0.609    &     0.000   &     17.605   &    1\\
        HD129502 &    0.38   &    -4.994 &       0.015    &   8.875   &    0.058 &      8.908   &    0.053     &   0.000    &    0.000     &   0.000  &      0.000   &    -0.280     &    0.000      &     inf   &    0\\

\noalign{\smallskip}
\hline

\end{tabular}
\tablebib{The observed P$_{rot}$ literature values are the median of available values from the following
  catalogues:
  \citet{Baliunas1996};  \citet{Wright2011}; \citet{Kiraga2012}; \citet{Armstrong2015}; 
  \citet{Samus2017}; \citet{Reinhold2020} \\

$^{a}$ The full table is provided at CDS. We show here the first ten rows as a guidance.
}
\normalsize
\end{sidewaystable}

\subsection{Correlation between \cahk\ and X-ray fluxes}\label{Sec:corr}

\begin{figure}
\begin{center}
\includegraphics[width=0.5\textwidth, clip]{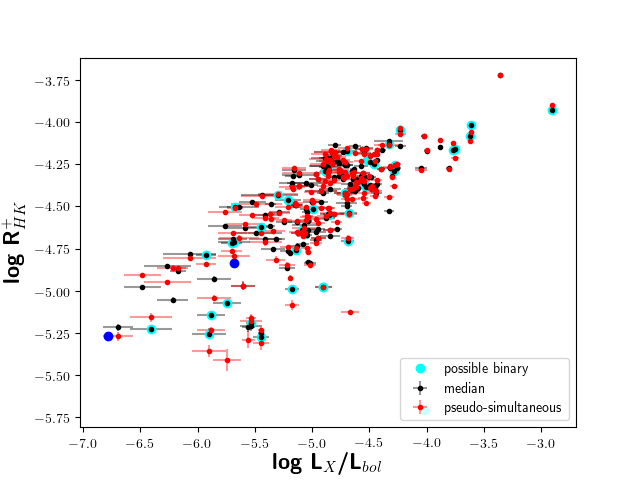}\\
\caption{\label{R+simul} Relation of \rplus to \lxlbol for all stars from the valid sample. Black 
dots show the median \rplus values derived from all available TIGRE spectra, and red dots show
\rplus values from the pseudo-simultaneous data. The blue points mark the values for the
Sun at solar maximum and solar minimum, respectively.
        }
\end{center}
\end{figure}

We first compared chromospheric to coronal activity, following the ansatz of
\citet{Mittag2018}, using \lxlbol as the coronal activity indicator and \rplus as a tracer of
chromospheric magnetic activity. 
In Fig.~\ref{R+simul} we
show the log(\rplus) values as a function of \lxlbol for the valid sample
as defined in Sec.~\ref{characterization}, first, using the median
of all available log(\rplus) values, and second, using the median of the pseudo-simultaneous values alone. For comparison, we also show the values for the Sun at solar
maximum and solar minimum, respectively. The X-ray luminosity was taken from
\citet{Judge2003} calculated for the ROSAT position-sensitive proportional counters (PSPC) energy band, which should be roughly comparable
with our values, see Sect. \ref{erosita}. The \rplus\ values were calculated from
indices taken from \citet{Schroeder2017} considering the minimum or maximum value for the
values they state for different cycles for solar minimum or maximum.
Additionally, we mark the 48 possible binaries we found during the
analysis, which mostly fall in the normal spread of the data. We inspected their properties
and found that only a minority
appears to be outliers, and we therefore incorporated them in our analysis
for the time being. We nevertheless
caution that some of the most active stars that flatten the correlation
may be binaries.

Formally, \lxlbol and \rplus are well correlated with values of $0.79$ and $0.77$ for
Pearson's correlation coefficient, $r$, if either all or only the pseudo-simultaneous data are
considered, with the p-value being $<10^{-35}$ in both cases. Fitting with a linear regression, we
find a gradient of 0.41 in both cases, which is in line with the well-known result that
the coronal X-ray flux is more sensitive to activity than the \ion{Ca}{ii} \rplus\, values.

\citet{Schrijver1983} first noted the necessity to correct the flux in the
\cahk\ lines for the photospheric and basal contributions. When \rp (which contains the basal flux) is used instead of \rplus, the correlation deteriorates, albeit only marginally.
The corresponding $r$-values for median log(\rp) of all data are 0.72 and also for the simultaneous data with
p-values $<10^{-30}$. Nevertheless, the slope of 0.3 resulting from the linear regression is lower
in both cases, and we attribute this to the relatively smaller contribution of the basal activity level
in the more magnetically active stars.  Because the relative basal flux contribution is clearly
strongest in the inactive stars, the slope of the relation is expected to flatten.
We therefore confirm that \rplus is more closely related to coronal activity and should therefore be preferred
over \rp in a comparison like this.

The appearance of magnetic activity is time dependent. Both geometrical effects such as
rotational modulation and changes in the magnetic field, for instance, attributable to magnetic cycles, contribute
to the variation. Consequently, a better relation between individual tracers may be expected in
data, which are sufficiently simultaneous to resolve these effects because the different layers
of the stellar atmosphere are expected to be in a comparable state.
However, no such behaviour is obvious in Fig.~\ref{R+simul} or in
the correlation coefficients stated above.

\begin{figure}
\begin{center}
\includegraphics[width=0.5\textwidth, clip]{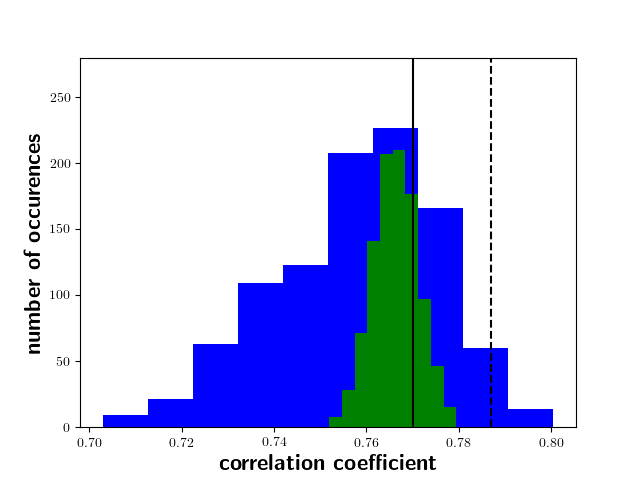}\\
\includegraphics[width=0.5\textwidth, clip]{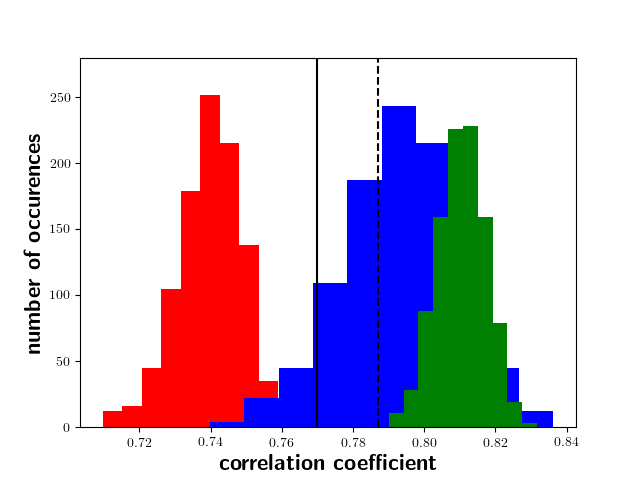}\\
\caption{\label{hist_corrcoeff} Correlation coefficients derived by choosing \rplus values
  from different spectra of the individual star and computing the correlation of \rplus
  to \lxlbol for all stars 1000 times. 
  \emph{Top:} Considering the valid sample and all available TIGRE spectra (blue), or only the
  pseudo-simultaneous data (green).
  \emph{Bottom:} Same as above, but for the 88 stars with TIGRE data older than 750 days
  before the start of the eROSITA observations. Choosing data only from these old
  data leads to a poorer correlation (red).
  For both panels, the vertical dashed black line represents the correlation coefficient
  obtained with the median value
  of  \rplus of all available data for the valid sample. The vertical solid 
  black line represents the same
  using pseudo-simultaneous data.
}
\end{center}
\end{figure}

To further investigate the case, we carried out a Monte Carlo simulation to study the distribution
of the correlation coefficients under the hypothesis that pseudo-simultaneity does not
improve the correlation between log(\rplus) and \lxlbol. To this end,
we obtained a series of realisations of the correlation coefficient between log(\rplus) and \lxlbol for
our program stars, opting for one arbitrary measurement of log(\rplus) for every star in the
computation. 
Similarly, we studied the distribution of the correlation coefficient considering only pseudo-simultaneous
data. In particular, we opted for one pseudo-simultaneously obtained pair
of log(\rplus) and \lxlbol for all stars and computed the correlation coefficient. We caution, however,
that for many sample stars, only one such pair is available (and only one 
eRASS1 \lxlbol measurement).
We executed this procedure 1000 times
and show the distribution of correlation coefficients in Fig.~\ref{hist_corrcoeff} (top panel).

As obvious from Fig.~\ref{hist_corrcoeff} , the distribution of correlation coefficients based on temporally uncorrelated data is far broader than
that based on pseudo-simultaneous pairs, which may partly be caused by the relative sparsity of these data.
The expected value derived from pseudo-simultaneous pairs is $0.76$, and the probability of obtaining a similar
or higher correlation coefficient from randomly associated pairs is about $20$\,\%.
Consequently, we find that the correlation based on pseudo-simultaneous pairs is not
significantly higher than that obtained from random associations. This is in line with
the  better correlation using the median of all measurements compared to
pseudo-simultaneous data. This comes as a surprise because when all measurements are used, an
intermediate chromospheric activity level is compared to individual coronal activity levels,
which may be associated with any state of activity.
Nevertheless, there is a slightly longer tail to poorer correlations in the distribution of the
temporally uncorrelated data.

Because non-coordinated X-ray to optical observation pairs are often separated by more than one year,
we repeated the simulations considering only TIGRE data, which are considerably ``older''
than the eROSITA data. This procedure obviously leads to a smaller stellar sample. For a time baseline between X-ray and
optical observation of at least 200 days, a second peak at lower correlation coefficients starts
to emerge. For a 750-day separation, 88 stars of the valid sample are left and a single peak
at a significant lower correlation coefficient than for the pseudo-simultaneous data, as is shown in the bottom panel
in Fig. \ref{hist_corrcoeff}.
This shift to a lower correlation for increasingly longer time baselines starting at about 200 days
suggests that cycles play
a larger role in the spread of the correlation of log(\rplus) and \lxlbol\, than rotation periods do.
This is in line with findings for HD~81809, for example, in which the chromospheric and coronal cycles correlate
well and L$_{\rm X}$ varies by nearly an order of magnitude \citep{Orlando2017}. Another example is 61~Cyg,
in which the
X-ray cycle is also analogue to the chromospheric cycle and X-ray brightness varies by a factor of three \citep{Robrade2012, Robrade2016}.

Nevertheless, there may be other reasons for the large spread that is also observed in the pseudo-simultaneous
data: Our pseudo-simultaneous data set is
hampered by the low number of data points per star. Statistical noise and possibly the effect of unresolved
short-term variability caused by flares both tend to reduce correlation.
Moreover, the pseudo-simultaneous window may be
too long for rapidly rotating stars, which may show a different part of their surface at the optical observations.
This is discussed in more detail in Sect. \ref{rotation}.


\subsection{Comparison to literature values}
\citet{Mittag2018} compared \rplus values obtained with TIGRE to non-simultaneous \lxlbol data
presented by \citet{Wright2011} in a sample of 169 main-sequence stars. 
The logarithms of \rplus and \lxlbol
exhibit a value of 0.76 for Spearmans's correlation coefficient and a slope of $0.40 \pm 0.02$,
which agrees well with the values we deduced. 

\citet{Schrijver1983} compared the X-ray surface flux and the excess flux
($\Delta$F) in the \cahk\ lines. The latter accounts for the photospheric and basal contributions
to the \cahk\ line. They used a power-law relation F$_{X} \sim \Delta$F$_{Ca}^{\beta}$ and
derived $\beta$ = 1.67. If we use surface fluxes instead of luminosities
\citep[using the interpolated
radius values from][]{Mamajek} for our data, we find $\beta$ = 1.60 with a
linear regression of the logarithmic values, which agrees well with the value of \citet{Schrijver1983}.

Another study to which we can compare our results is the one by \citet{MartinezArnaiz2011}, who used
observations of
298 F- to M-type stars for constructing various flux--flux relationships. Because the authors did not directly
compare log F$_{\cahk}$ and log F$_{X}$, we used the bluest component of the \cairt and
their log F$_{Ca II 8498}$ to log F$_{\cahk}$ conversion from \citet{MartinezArnaiz2011err}
additionally. Based on this, we deduce
a slope of 1.58 from their studies, which is within a few percent of our number and that of
\citet{Schrijver1983} again.

\subsection{Rotation and the pseudo-simultaneity}\label{rotation}

The duration of our time window for pseudo-simultaneous observations
of three days before and after the start and end of the X-ray observations
is largely determined by observational constraints.
For stars with short rotation periods, rotational modulation can be relevant
on this timescale.
Evidently, a star with a rotation period of 10~days switches visible
hemispheres every 5~days and
rotates by a quarter in $2.5$~days, where the latter already meets our
criterion of pseudo-simultaneity.

To investigate the effect on our analysis, we
first assessed the frequency of strong chromospheric variations on short
timescales.
For the 183 valid sample stars, we have a total of about 5900 TIGRE spectra,
2100 of which show a time difference of less than three days
to the following spectrum. Out of these high-cadence pairs, 80
show a variation of more than 0.1
in log(\rplus), which belong to 39 stars. This variation is about 100 times
higher than the mean variation between spectra for the valid sample.
While underestimated errors may play a role for some stars, there
are many time series for which this is clearly not the case.
One star (the  late-F type planet host star
HD~120136/$\tau$ Boo; P$_{ro}$=$3.1$~d by \citet{Mittag2017}) shows the largest number of detected short-term
variations by far and seems to be peculiar to this respect.
Out of its 780 TIGRE spectra, 25 fulfil our
short-term variation criterion with many variations even
occurring within a single day.
The ultimate reasons for these strong variations remain unknown, but because these variations are usually relatively seldom, we argue that for most stars,
they are not
caused by rotation,
but by flares.

To place the length of the pseudo-simultaneous observational window in the
context of the rotational timescale,
we need to know the stellar rotation periods.
Unfortunately, directly measured rotation
periods are available for only 99 sample stars, 62 of which belong to the
valid sample.
Fast-rotating stars tend to be more active and their periods easier to measure, 
and, indeed,
41 out of the 62 periods in the valid sample are shorter than 10 days; see for instance \cite{Schmitt2020}
for the difficulties in measuring accurate periods in low-activity stars.

\begin{figure}
\begin{center}
\includegraphics[width=0.5\textwidth, clip]{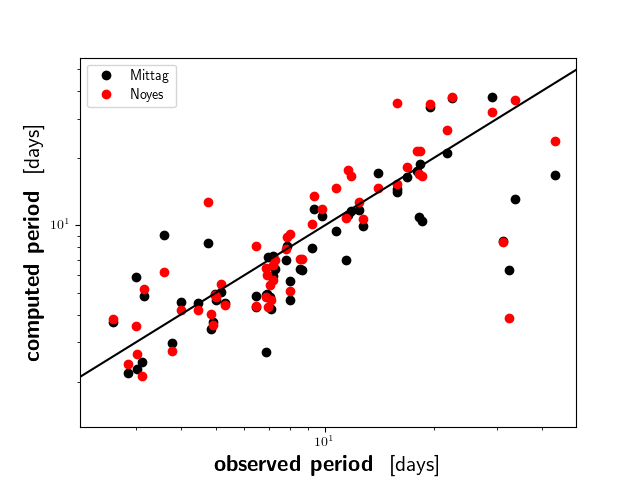}\\
\caption{\label{prot} Computed P$_{rot}$ vs. observed P$_{rot}$ for 59 stars of the valid sample with measured P$_{rot}>2.$ days.}  
\end{center}
\end{figure}

As an alternative to a direct measurement, rotation periods can be estimated
based on the
stellar activity level. Both \citet{Noyes1984} and
\citet{Mittag2018} provided estimates of the rotation period,
P$_{\mathrm{rot}}$, based on \rplus.
Table~\ref{measurements} gives the estimated rotation period based on the
relation by
\citet{Mittag2018} along with an observed literature period whenever available.
A comparison of observed and calculated
rotation period is shown in Fig.~\ref{prot}, where we consider both the
estimates based on the
\citet{Noyes1984} (red points) and \citet{Mittag2018} relations (black
points). Except for
some outliers, the activity-based estimates and the observed rotation
periods are
typically consistent to within about 3 days.
The largest discrepancies occur among inactive stars, where the
estimate
of the basal flux also has the strongest effect on the actual \rplus\, value.

However, it is not necessarily the period estimate that is in error.
Period searches are known to be sensitive also to
multiples of the correct period. Cases in point may be the two
stars with a measured period of about 40~d and a computed period of about
20~d (see Fig.~\ref{prot}),
for which the latter, shorter-period estimate appears more plausible in
terms of activity.

The valid sample contains 80 estimated rotation periods longer than 10
days and 103 rotation periods shorter than 10 days.
As rotational modulation may therefore play a role in our consideration of
pseudo-simultaneity,
we divided the valid sample into slow and fast rotators with estimated
rotation periods above and below 10 days, respectively.

For the fast rotators, the correlation coefficient between \rplus\ and
\lxlbol\ is 0.75 considering the median
\rplus\, values and 0.76 for the pseudo-simultaneous \rplus\, values. The
corresponding coefficients for the slow
rotators are 0.72 and 0.70. The slopes are 0.39 for the fast and 0.43
for the slow rotators, with little dependence on whether pseudo-simultaneous
or median values are considered.
The pseudo-simultaneity apparently plays a smaller
role than errors introduced by measurements of line indices or their
conversion for inactive stars in our current
data set. More pseudo-simultaneous measurements in the context of the
ongoing eROSITA survey will us allow to
investigate this issue in more detail.

\subsection{Prediction of \lxlbol\, from \rplus\, values}\label{secspectype}
Applying a linear regression to our measured \lxlbol\ and \rplus\
values, we obtain best-fit values for the coefficients $a$ and $b$
in the relation
\begin{equation}
    \mbox{\lxlbol} = a \times \log(\mbox{\rplus}) + b \;.
    \label{eq:R}
\end{equation}
The values obtained for the entire valid sample as well as for spectral-type selected
sub-samples are given in Table~\ref{spectype} and shown in
Appendix~\ref{appspectype}. For M~dwarfs we do not have enough data points
to define a meaningful fit. 

\begin{figure}
\begin{center}
\includegraphics[width=0.5\textwidth, clip]{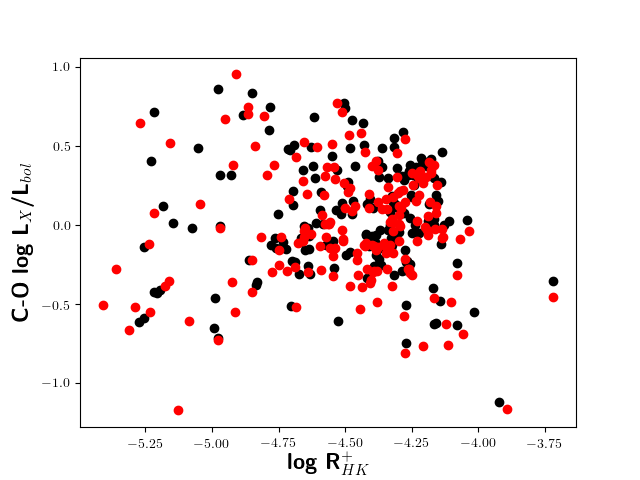}\\
\caption{\label{prediction} Computed minus observed \lxlbol values as a function of
measured P$_{rot}$ for stars of the valid sample with available P$_{rot}$.
 The black dots represent the median data, and the red dots represent 
the pseudo-simultaneous data.
}
\end{center}
\end{figure}

The relation in Eq.~\ref{eq:R} can be used to predict \lxlbol\ from \rplus, the measurement of which requires
no space-based instrumentation.
In Fig.~\ref{prediction} we show the comparison of the calculated and observed \lxlbol\ values as a function
of \rplus, that is, the residuals with respect to the best-fit relation.
The standard deviation of the pseudo-simultaneous data in calculated minus 
observed (C-O) \lxlbol is 0.36 and 0.35 for the median data for the valid sample.
Overall, it appears that \lxlbol\ is under-predicted for the most active stars,
which also tend to be fast rotators. The reason for this under-prediction
  remains unknown. A possible physical reason would be the phenomenon of
super-saturation, which is observed
  in X-rays \citep[][e.\,g.]{Thiemann2020} but not in chromospheric lines \citep[][e.\,g.]{Christian2011}.
However, we do not consider this very likely because the stars in our valid sample are not active enough
to fall in the regime of super-saturation for X-rays, which starts for stars with \vsini $>$ 100\,km\,s$^{-1}$.
Only very few of our stars are even located at around \lxlbol=-3.14 as was observed for the saturated
regime by \citet{Thiemann2020}. There are two additional explanations:
(i) these most active stars can be fast rotators that may cause a leaking of chromospheric flux out of the
wavelength range considered for the integration for the index calculation. We
are aware of two stars in the valid sample with a \vsini\, higher than
40~km\,s$^{-1}$.
(ii) Three of the binary candidates fall into this regime.

\begin{figure}
\begin{center}
\includegraphics[width=0.5\textwidth, clip]{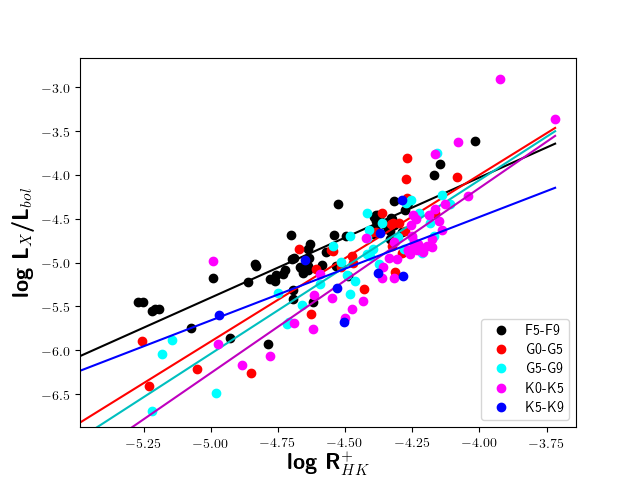}\\
\caption{\label{linregresstype} \lxlbol\, values as a function of \rplus\, for different
spectral types as defined in Sect. \ref{secspectype}. The F and late-K type stars have
the lowest gradient in the linear regression fit.
}
\end{center}
\end{figure}

Inspecting the coefficients for the
spectral subgroups defined in Table~\ref{spectype}, we find that
the gradients for the F and late-K type stars are lower
than for the other stars. For the F-type sub-sample, this is caused by a group of
stars with the lowest \rplus\ values (cf. Fig.~\ref{linregresstype}).
There is also a slight vertical
offset between the F spectral type group and the other groups for low to moderate activity levels.
For the lowest activity stars, even small errors in the estimation of the
basal flux (resulting, e.g. from poor fits of the PHOENIX models) lead to larger 
errors on the calculated \rplus\ values than is the case for more active stars, for which 
the magnetic contribution to the line flux is much higher than the basal flux level. Therefore
we repeated the fits, omitting the most inactive stars with $\log(\mbox{\rplus}) < -4.9$.
This led to higher values for the slopes in all spectral subgroups with a
significant increase observed for the F-type sub-sample.
Using the spectral-type dependent relations (all valid sample stars) to predict
\lxlbol leads to a slightly smaller standard deviation of 0.29 in C-O \lxlbol,
but \lxlbol\ remains under-predicted in the most active stars.

\begin{table}
  \caption{\label{spectype} Best-fit coefficients for Eq.~\ref{eq:R} with standard error.}
\footnotesize
\begin{tabular}[h!]{lcc}
\hline
\hline
\noalign{\smallskip}

Spectral type   & $a$ & $b$ \\

\noalign{\smallskip}
\hline
\noalign{\smallskip}
\multicolumn{3}{c}{All stars from valid sample (\rplus)}\\
All   & 1.51 $\pm$ 0.08 & 1.90 $\pm$ 0.35\\
F5-F9 & 1.37 $\pm$ 0.06 & 1.45 $\pm$ 0.20\\
G0-G4 & 1.90 $\pm$ 0.14 & 3.60 $\pm$ 0.54\\
G5-G9 & 1.98 $\pm$ 0.08 & 3.86 $\pm$ 0.31\\
K0-K4 & 2.11 $\pm$ 0.13 & 4.29 $\pm$ 0.46\\
K5-K9 & 1.18 $\pm$ 0.05 & 0.24 $\pm$ 0.20\\
\hline
\multicolumn{3}{c}{Valid sample with most inactive stars excluded}\\
F5-F9 & 1.72 $\pm$ 0.06 & 2.98 $\pm$ 0.24\\
G0-G4 & 2.25 $\pm$ 0.20 & 5.08 $\pm$ 0.82\\
G5-G9 & 2.06 $\pm$ 0.09 & 4.23 $\pm$ 0.33\\
K0-K4 & 2.65 $\pm$ 0.13 & 6.55 $\pm$ 0.50\\
K5-K9 & 1.45 $\pm$ 0.05 & 1.43 $\pm$ 0.21\\
\hline
\multicolumn{3}{c}{All stars from valid sample (\rp$_{\rm Noyes}$)}\\
All   & 2.31 $\pm$ 0.11 & 5.71 $\pm$ 0.46\\
F5-F9 & 2.29 $\pm$ 0.12 & 5.71 $\pm$ 0.49\\
G0-G4 & 2.85 $\pm$ 0.22 & 8.04 $\pm$ 0.92\\
G5-G9 & 2.84 $\pm$ 0.14 & 8.00 $\pm$ 0.54\\
K0-K4 & 2.77 $\pm$ 0.17 & 7.75 $\pm$ 0.70\\
K5-K9 & 2.11 $\pm$ 0.09 & 4.86 $\pm$ 0.34\\
\noalign{\smallskip}
\hline

\end{tabular}
\normalsize
\end{table}


For comparison, we also used \rp\, in the linear regression because the relation
by \citet{Noyes1984} remains widely used. We give here coefficients for the
relation equivalent to Eq.~\ref{eq:R} as well,
\begin{equation}
    \mbox{\lxlbol} = a \times \log(\mbox{\rp}_{\rm Noyes}) + b\; ,
\end{equation}
which leads to a standard deviation of 0.34 for C-O \lxlbol
values, comparable to the value obtained with \rplus. 

Finally, we can predict  \lxlbol for the stars in our sample without
eROSITA detections. By definition, these do not belong in our valid
sample. All but one of them have \cahk\, index measurements by TIGRE,
and none has a B$-$V < 0.44~mag, which would place it outside the range of
definition of our conversion relations. The calculated \lxlbol\, values for these stars
are lower than those for the valid sample on average. While the median value of \lxlbol
of the valid sample is $-4.86$, the predicted median of the stars without eROSITA X-ray
detection is $-5.99$. This places most of these stars at the low-activity end of our distribution. While we cannot check the prediction at the moment, it is
at least consistent with the current lack of X-ray detections.

\section{Conclusions}

We investigated the chromospheric coronal connection for a sample of 404 magnetically active
stars using simultaneously measured X-ray fluxes from the eRASS1 survey as tracer 
of the coronal activity 
and \cahk\ indices derived from spectra obtained by the TIGRE telescope. From these
indices, we derived \rplus values, which we used as a proxy
for the chromospheric activity state. The sample we presented here of (pseudo-)simultaneous 
coronal and chromospheric measurements is to our best knowledge the largest such sample obtained to date, comprising 183 mid-F to early-M dwarfs.

The correlation between chromospheric activity
measured through log(\rplus)  and coronal activity measured through \lxlbol\, somewhat surprisingly is almost identical for the  median
and the simultaneously measured \rplus values.  This finding is corroborated by a
bootstrap analysis, in which we randomly assigned a chromospheric measurement to the X-ray measurement.
On the other hand, if only optical data ``older'' than two years compared to the eROSITA X-ray data are 
used in the bootstrap analysis, the random data pairs lead to a lower correlation than the pseudo-simultaneous data.
Thus, the pseudo-simultaneous data or time series improve the
correlation and are therefore desirable. In particular,  the effect of activity variations on longer timescales (e.g. cycles) seems to be more important than short-term effects (e.g. rotational modulation).
The effect of using median \lxlbol\, values can only be investigated in
the future, when more eROSITA measurements with pseudo-simultaneous TIGRE
data become available.  We also stress that we considered only sample correlations because eROSITA time
series are not yet available.

Computing the correlation between log(\rplus) and \lxlbol\, for different
sub-samples in spectral type shows a tendency to higher slopes for earlier
type stars. This may indicate a stronger
coupling of the chromosphere and corona in earlier-type stars, but it may also
be caused by an insufficient correction for the effect of spectral type by
the \rplus\, conversion. Measuring the \cahk\, index with high enough precision
and correcting for the basal flux seems to be an issue for the most inactive
stars where even slightly incorrect estimates of the basal flux have a relatively strong effect on the \rplus\ values and where statistical noise also has a relatively stronger effect on these weak lines. 
Accordingly, we find the worst correlation between log(\rplus) and \lxlbol\,
for the pseudo-simultaneous data of the slow rotators, where the opposite is
expected based on timing considerations.

Predicting \lxlbol\, from \rplus\, values is very much appealing because X-ray measurements are
far more difficult to come by than ground-based determinations of the \ion{Ca}{ii} H and K line flux.
The predictions seem to work well for many
stars with an expected standard deviation of about 0.34. We also predict
mostly low \lxlbol\, values for the stars without an eROSITA
detection.
We conclude that the correlation between \lxlbol\, and \rplus\,
is suitable to make predictions of \lxlbol\, from measured  
optical data accurate to within factors of a few.
While the origin of the considerable
scatter in the relation remains basically unknown, we expect contributions from (i) shortcomings
in the conversion into \rplus\, values (and in the basal flux estimation) or
(ii) in the non-simultaneity of the pseudo-simultaneous data for the fast
rotators, and (iii) in an uncoupling of the chromospheric and coronal tracers
even after subtraction of the non-magnetic heating for the chromosphere.
Our results
demonstrate that the ongoing eROSITA X-ray measurements in the framework of the all-sky survey along
with further pseudo-simultaneous measurements by TIGRE are likely to significantly improve our understanding
of the chromospheric coronal connection.

\begin{acknowledgements}
  B.~F. acknowledges funding by the DFG under Schm \mbox{1032/69-1}.

  This work is based on data from eROSITA, the soft X-ray instrument aboard SRG, a joint
  Russian-German science mission supported by the Russian Space Agency (Roskosmos), in the
  interests of the Russian Academy of Sciences represented by its Space Research Institute
  (IKI), and the Deutsches Zentrum f\"ur Luft- und Raumfahrt (DLR). The SRG spacecraft was
  built by Lavochkin Association (NPOL) and its subcontractors, and is operated by NPOL
  with support from the Max Planck Institute for Extraterrestrial Physics (MPE).

  The development and construction of the eROSITA X-ray instrument was led by MPE, with
  contributions from the Dr. Karl Remeis Observatory Bamberg \& ECAP (FAU
  Erlangen-Nuernberg), the University of Hamburg Observatory, the Leibniz Institute for
  Astrophysics Potsdam (AIP), and the Institute for Astronomy and Astrophysics of the
  University of T\"ubingen, with the support of DLR and the Max Planck Society. 

This work is based on data obtained with the TIGRE telescope, located at
La Luz observatory, Mexico. TIGRE is a collaboration of the Hamburger
Sternwarte,
the Universities of Hamburg, Guanajuato and Li\`ege.

This research has made use of the SIMBAD database, operated at CDS, Strasbourg, France.

\end{acknowledgements}

\bibliographystyle{aa}
\bibliography{papers}

\appendix
\section{\lxlbol vs. \rplus per spectral type}\label{appspectype}

\begin{figure*}
\begin{center}
\includegraphics[width=0.5\textwidth, clip]{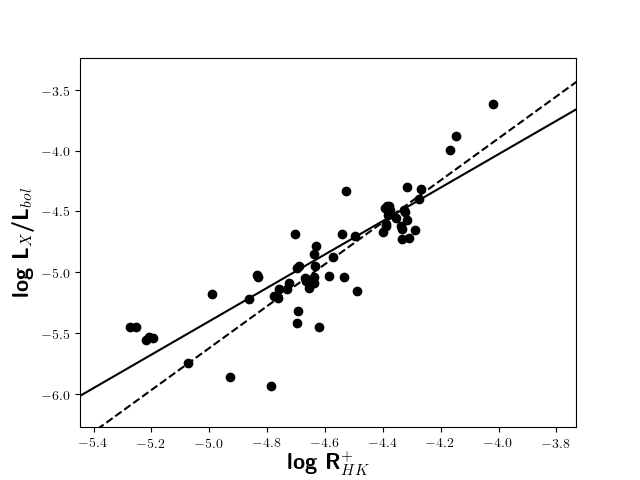}
\includegraphics[width=0.5\textwidth, clip]{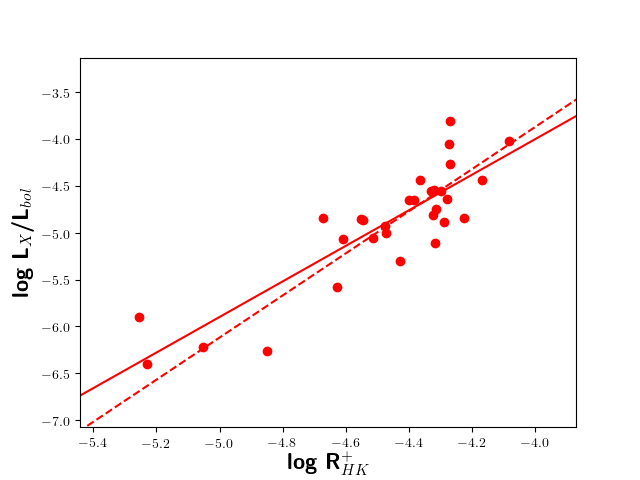}\\
\includegraphics[width=0.5\textwidth, clip]{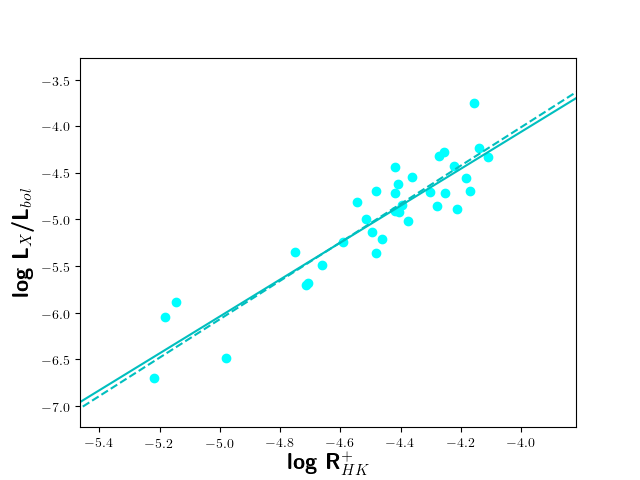}
\includegraphics[width=0.5\textwidth, clip]{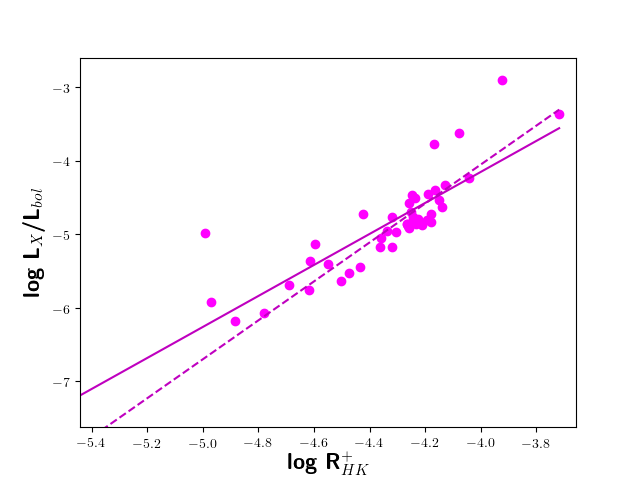}\\
\includegraphics[width=0.5\textwidth, clip]{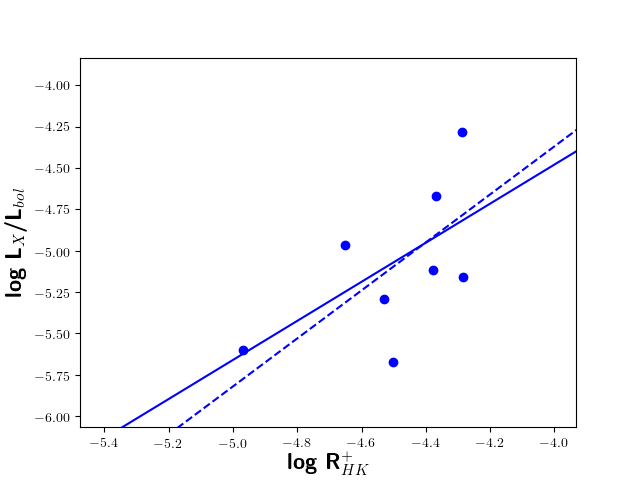}
\includegraphics[width=0.5\textwidth, clip]{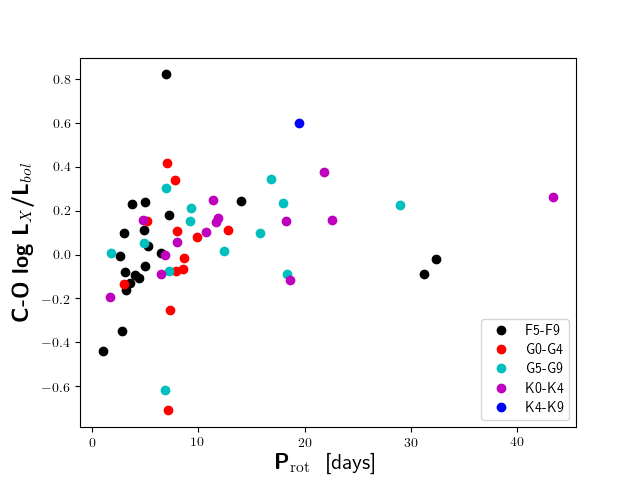}
\caption{\label{R+spectype} Same as Fig. \ref{linregresstype}, but for every spectral type group separately. The solid lines mark the linear regression taking all stars into account, andthe dashed lines mark the linear regression with
only the more active stars with log(\rplus) $>$ -4.9. In the bottom right corner, we show the
same as in Fig. \ref{prediction}, but for the fits of the different spectral groups
marked in their respective colour.
}
\end{center}
\end{figure*}

We show  the best fits for the different spectral groups
together with the data in Fig. \ref{R+spectype}.
The obtained linear best fits are marked for the whole valid sample by a solid line
and excluding the
least active stars by a dashed line.

 




\end{document}